\documentclass[%
 prl,twocolumn,
 nofootinbib,
groupedaddress,
 amsmath,amssymb,
 aps,
floatfix
]{revtex4-2}
\usepackage{amsmath}
\usepackage{graphicx,xcolor}
\usepackage{dcolumn}
\usepackage{bm}
\usepackage{hyperref}

\begin{document}


\title{Observing gravitational waves with solar system astrometry}

\author{Giorgio Mentasti}%
\author{Carlo R. Contaldi}
\affiliation{%
 Blackett Laboratory, Imperial College London, SW7 2AZ, UK}

\date{\today}

\begin{abstract}
The subtle influence of gravitational waves on the apparent positioning of celestial bodies offers novel observational windows \cite{Braginsky:1989pv,Pyne:1995iy,1993ApJ...418..202F,Kaiser:1996wk}. We calculate the expected astrometric signal induced by an isotropic Stochastic Gravitational Wave Background (SGWB) in the short distance limit. Our focus is on the resultant proper motion of Solar System objects, a signal on the same time scales addressed by Pulsar Timing Arrays (PTA). We derive the corresponding astrometric deflection patterns, finding that they manifest as distinctive dipole and quadrupole correlations or, in some cases, may not be present. Our analysis encompasses both Einsteinian and non-Einsteinian polarisations. We estimate the upper limits for the amplitude of SGWBs that could be obtained by tracking the proper motions of large numbers of solar system objects such as asteroids. We find that for SGWBs with negative spectral indices, such as that generated by Super Massive Black Hole Binaries (SMBHB), the constraints from these observations could rival those from PTAs. With the Gaia satellite and the Vera C. Rubin Observatory poised to track an extensive sample of asteroids—ranging from ${\cal O}(10^5)$ to ${\cal O}(10^6)$, we highlight the significant future potential for similar surveys to contribute to our understanding of the SGWB.
\end{abstract}

\pacs{Valid PACS appear here}
\maketitle

{\sl Introduction.--} The detection of Gravitational Waves (GWs) by ground-based observatories \cite{PhysRevLett.116.061102} has opened a new window onto the universe. So far, observations are probing waves at around 100 Hz produced during the merger of compact binaries. The LISA space-based mission will probe a window around $10^{-3}$ Hz \cite{2019BAAS...51g..77T} that is rich in expected galactic and extragalactic signals. Yet, another frontier, focusing on low-frequency gravitational waves, at around $10^{-8}$ Hz, is probed by Pulsar Timing Arrays (PTAs) with recent exciting hints of detection of a GW background \cite{EPTA:2023fyk,NANOGrav:2023gor,refId0}. This window is also potentially probed by astrometric measurements \cite{2011PhRvD..83b4024B,Darling2018}.

PTAs employ a network of monitored millisecond pulsars dispersed across our galaxy. These pulsars act as precise cosmic clocks. Gravitational waves, when passing through the galaxy, cause tiny changes in the observed times of arrival of pulses from these distant sources. A concerted effort by a multitude of observatories worldwide aims to observe this effect, hunting for the signature of, for instance, supermassive black hole binaries which are expected to produce gravitational waves in the PTA-sensitive frequency range.

On the other hand, astrometric measurements, obtained through highly precise observations of the positions and motions of stars, can also be influenced by gravitational waves. The waves can induce apparent shifts in the position of stars on the celestial sphere, offering another method of detection. With the advent of missions like Gaia \cite{Gaia}, astrometry has reached an unprecedented precision, allowing us to explore the gravitational wave universe in new and complementary ways.

PTAs and astrometric measurements are pioneering the exploration of the low-frequency gravitational wave spectrum. Detection of signals in this frequency range offers the potential to increase our understanding of massive astrophysical processes and challenge our theories of gravity. Both these measurements have been considered in the distant source limit because the distances between the Earth and the objects being tracked, of ${\cal O}(10^3)$ pc, are much longer than the wavelength of the GWs being probed, typically of ${\cal O}(1)$ pc. In this limit, the timing and astrometric distortions are dominated by the effect of GWs effects at the observer's position on the Earth. The effects due to GWs at the position of the objects are assumed to be uncorrelated and do not contribute to the signal, only its variance.

In this {\sl letter} we consider the opposite, short distance limit of astrometric observations. This is useful when considering solar system bodies as the objects being tracked on timescales of years. In this case, the distances involved are on the order of an AU and the GW wavelengths of interest are ${\cal O}(10^6)$ AU. In the short distance limit, the distortion induced by GWs is correlated at the observer's and the source's position. This results in a different form for the signal. Of particular interest is the form of the correlation in the astrometric signal between pairs of sources, the analogue of the Hellings-Down (HD) curve \cite{HD}. The correlations encode the symmetries of the polarisation of the underlying GWs and can provide a `smoking gun' for the nature of the signal. For example, further increases in the significance of the recent hint of the detection of an HD correlation in PTA signals will ultimately confirm the underlying tensorial nature of the signal.

At first, it may seem strange to consider this limit, particularly for astrometric observations that necessitate tracking large numbers of objects. However, current surveys, along with ones planned in the near future, are already operating at scales that make it worthwhile to explore this possibility. The Gaia satellite's primary goal is to track the apparent position of billions of galactic stars but it also observes ${\cal O}(10^5)$ asteroids in the solar system with micro-arcsecond (mas) astrometric accuracy \cite{Gaia_asteroids}. The size of the detected catalogue of asteroids and its astrometric accuracy is limited by sensitivity since asteroids are dimmer than stars. Soon, the Vera C. Rubin observatory, currently being completed in Chile, will return astrometric tracking data of ${\cal O}(10^6)$ asteroidal objects with an accuracy of around 50 mas \cite{2019ApJ...873..111I} with a cadence of a few days and over a large fraction of the sky. As we argue below, these observations may provide a new opportunity for measuring low-frequency GWs.

{\sl Formalism.--} We consider a Stochastic GW Background (SGWB) as a superposition of plane waves. After a suitable choice of gauge, this can be written as
\begin{align}\label{planar_wave}
	h_{ij}(t,\mathbf{x})=\int df     \int d^2\mathbf{p}\sum_\lambda h^\lambda(f,\mathbf{p})e^\lambda_{ij}(\mathbf{p})e^{-2\pi if(t-\mathbf{p}\cdot\mathbf{x})}\,,
\end{align}
where $i$, $j$, etc. run over spatial coordinates and the index $\lambda$ runs over all orthogonal polarisation state including.  We include the standard Einsteinian, transverse, traceless polarisations $+$ and $\times$ and non-Einsteinian scalar ($S$), vectorial ($X$ and $Y$), and longitudinal ($L$) polarisations \cite{PhysRevD.97.124058}.
In \eqref{planar_wave}, we considered only planar waves with a standard dispersion relation propagating at the speed of light. However, the generalisation to subluminal non-Einstanian modes \cite{PhysRevD.103.024045,PhysRevD.107.044007} would not be cumbersome, given the short-distance limit we are considering.
The unit vector $\mathbf{p}$ is aligned with the momentum of each wave with frequency $f$.

The amplitude of $h^\lambda(f,\mathbf{p})$ in \eqref{planar_wave} is a stochastic variable at each frequency $f$ and direction $\mathbf{p}$. We assume the SGWB to be stationary, isotropic, and Gaussian such that
\begin{align}\label{PSD_signal}
	\langle h^\lambda(f,\mathbf{p})h^{\lambda\star}(f',\mathbf{p'})\rangle=\delta_{\lambda\lambda'}\delta(f-f')\delta^{(2)}(\mathbf{p}-\mathbf{p'})H(f)\,.
\end{align}
The astrometric deflection induced by gravitational waves has been studied in (see e.g. \cite{PhysRevD.101.024038,PhysRevLett.119.261102,PhysRevD.97.124058,2012A&A...547A..59M,Qin2019,GOLAT2021136381}). The deflection of the apparent position of a source on the sky is denoted by a two-dimensional vector denoting the perturbation of the true direction unit vector of the source $\mathbf{n} \to \mathbf{n}+\mathbf{\delta n}$. In the limit where the distance between the observer and the source object is much smaller than the wavelength of the gravitational waves, the expression for the astrometric deflection components (see, e.g. equation (19) of \cite{PhysRevD.97.124058}) simplifies to
\begin{align}\label{dn}
	\mathbf{\delta n}^{i}(\mathbf{n},t)&=\frac 1 2 \int df\int d^2\mathbf{p}\sum_\lambda h^\lambda(f,\mathbf{p})e^\lambda_{jk}(\mathbf{p}) \mathrm{e}^{-2\pi if t}\times\nonumber\\
	& \mathbf{n}^j\left(\delta^{ik}-\mathbf{n}^i\mathbf{n}^k\right)\,.
\end{align}
This expression is valid in the limit we consider here. For example, if we consider the typical distance to candidate objects in the solar system at a distance $d_a$ of order ${\cal O} (1)$ AU from an observer at the Earth's location, and if the proper motion of such objects were tracked on timescales $T$ of ${\cal O}(10)$ years, the scaling $2\pi d_a/(cT)\simeq 1\times 10^{-5}$ that appears in e.g. (19) of \cite{PhysRevD.97.124058} is very small and one can perform a Taylor expansion in that quantity.

The unequal-time correlator between an object with line-of-sight $\mathbf{n}$ and one with line-of-sight $\mathbf{n}'$
becomes
\begin{align}\label{dndn}
	\langle\delta n^{i}(\mathbf{n},t)\delta &n^{j}(\mathbf{n'},t')\rangle=\frac 1 4 \int df H(f)e^{2\pi if(t'-t)}\times\nonumber\\
	&\Gamma_{abcd}\mathbf{n}^a \,\mathbf{n'}^c\left(\delta^{ib}-\mathbf{n}^i\mathbf{n}^b\right)\left(\delta^{jd}-\mathbf{n'}^{j}\mathbf{n'}^{d}\right)\,,
\end{align}
where the tensor $\Gamma_{abcd}$ is defined as
\begin{align}\label{Gamma}
	\Gamma_{abcd}\equiv\int d^2\mathbf{p}\sum_{\lambda} e^\lambda_{ab}(\mathbf{p}) \,e^{\lambda}_{cd} (\mathbf{p})\,,
\end{align}
where we adopted the convention of \cite{PhysRevD.97.124058} for the polarisation tensors.
The contributions to the $\Gamma$ can be separated into Einsteinian transverse traceless $(++)$ + $(\times\times)$ ($E$), and non-Einsteinian, scalar transverse ($S$), vectorial $(XX)$ + $(YY)$ ($V$), and longitudinal ($L$) to give
\begin{align}
	\Gamma_{abcd}^{E}&=
	&\frac{8\pi}{15}\left[3(\delta_{ac}\delta_{bd}+\delta_{ad}\delta_{bc})-2 \delta_{ab}\delta_{cd}\right]\,,\label{GammaGR}\\
	\Gamma_{abcd}^{S}&=
	&\frac{4\pi}{15}\left[(\delta_{ac}\delta_{bd}+\delta_{ad}\delta_{bc})+6\delta_{ab}\delta_{cd}\right]\,,\\
	\Gamma_{abcd}^{V}&=
	&\frac{8\pi}{15}\left[3(\delta_{ac}\delta_{bd}+\delta_{ad}\delta_{bc})-2 \delta_{ab}\delta_{cd}\right]\,,\\
	\Gamma_{abcd}^{L}&=
	&\frac{8\pi}{15}\left[\delta_{ac}\delta_{bd}+\delta_{ad}\delta_{bc}+\delta_{ab}\delta_{cd}\right]\,.\label{GammaL}
\end{align}
We now consider equal-time cross-correlations (i.e. setting $t'=t$ in \eqref{dndn}), for each pair of objects and each observed time in a reference frame where the first object is aligned with the cartesian $\mathbf{z}$-axis and the second has no cartesian $\mathbf{y}$ component such that $\mathbf{n}=\mathbf{z}$, $\mathbf{n}'=\mathbf{x}\sin\theta+\mathbf{z}\cos\theta$ and $\mathbf{n}\cdot \mathbf{n}'= \cos \theta$.
The results will be independent of this rotation for an isotropic SGWB being considered here. We define the projected angular deflections similarly to \cite{PhysRevD.97.124058}.
\begin{align}\label{dphi_dtheta_def}
	\delta\phi(\mathbf{n})&=\mathbf{y}\cdot\mathbf{\delta n}(\mathbf{n})\nonumber\\
	\delta\theta(\mathbf{n})&=(\mathbf{x}\cos\theta-\mathbf{z}\sin\theta)\cdot\mathbf{\delta n}(\mathbf{n})\,.
\end{align}
Giving four possible correlations $\langle\delta\phi\delta\phi\rangle$, $\langle\delta\theta\delta\theta\rangle$, $\langle\delta\theta\delta\phi\rangle$, and $\langle\delta\phi\delta\theta\rangle$. These can be calculated by carrying out the contraction in \eqref{dndn} using the definitions \eqref{GammaGR}-\eqref{GammaL}. As expected from symmetry arguments, we confirm that all cross terms vanish for any polarisations. The two diagonal correlations are summarised in Table~\ref{tab:correlators} for all polarisations where
\begin{align}
	\left(\!\!\begin{array}{cc} \langle\delta\phi\delta\phi\rangle &\langle\delta\phi\delta\theta\rangle \\
		\langle\delta\theta\delta\phi\rangle &\langle\delta\theta\delta\theta\rangle \end{array}\!\!\right) &=\frac{2\pi}{5}\left(\!\!\begin{array}{cc} \xi_{\phi\phi} & 0 \\
		0 &\xi_{\theta\theta} \end{array}\!\!\right)\int df\, H(f)\,.
\end{align}

\begin{table}[t!]
	\centering
	\caption{\label{tab:correlators}Correlation functions}
	\begin{tabular}{l c c c c}
		\hline
		& E & S & V & L \\
		\hline
		$\xi_{\phi\phi}(\theta)$ & $\cos(\theta)$ & $\frac{1}{6}\cos(\theta)$ & $\cos(\theta)$ & $\frac{1}{3}\cos(\theta)$\\[3pt]
		$\xi_{\theta\theta}(\theta)$ & $\cos(2\theta)$ &$\frac{1}{6}\cos(2\theta)$ & $\cos(2\theta)$ & $\frac{1}{3}\cos(2\theta)$\\[1pt]
		\hline
	\end{tabular}
	\begin{center}
		\footnotesize
		\textbf{Note:} The angular dependence of the correlation functions of the projected deflections. For all polarizations, the $\delta\phi$ correlator is a dipole, and the $\delta\theta$ correlator is a quadrupole; only the relative normalizations differ. The off-diagonal correlators all vanish. The correlation functions are the analogs of the Hellings-Down curve for PTAs. We extracted a common factor of $\frac{2\pi}{5}\int df \,H(f)$
	\end{center}
\end{table}

\begin{figure}[!t]
	\centering
	\includegraphics[width=\linewidth,clip]{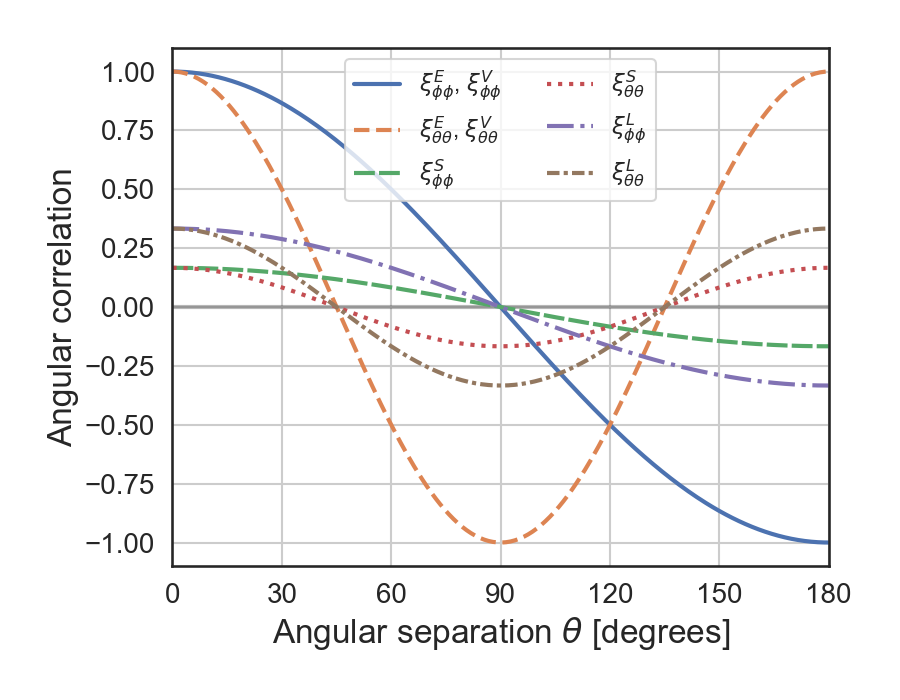}
	\caption{The GW induced angular correlations $\xi(\theta)$ as a function of the angular separation of tracked object pairs $\theta$. These functions are analogous to the Hellings-Down curves for PTA observations. All correlations are either purely dipolar or quadrupolar and are degenerate in all polarisations, up to an overall factor. 
	}
	\label{fig:correlations}
\end{figure}

These correlations are simpler in form than the Hellings-Down curve for PTAs and its analogues for astrometry \cite{PhysRevD.97.124058,GOLAT2021136381} in the distant source limit. This is because, in the short distance limit, the absence of additional phases in the projection of the polarisations simplifies their angular integration that is involved in the calculation, whilst their symmetries are also fully preserved and remain manifest in the rotated frame that defines the deflection $(\delta\theta, \delta\phi)$.
Interestingly, we note that an exact quadrupolar pattern, as in figure~\ref{fig:correlations}, appears for the case of tensorial modes from spin-2 dark
matter \cite{2024arXiv240203984C}. This is one of the possible cases where you have a tensorial pattern that differs from the Hellings-Downs signature. Many other examples are present in literature \cite{Wang:2023div,PhysRevD.107.L101502,2023arXiv231007537B} for both luminal and subluminal gravitational waves.
Another point to consider is that the correlators of the angular aberrations sourced by vectorial modes do not show the divergence for a relative angle of $\xi=0$, which is present in the long-distance limit \cite{PhysRevD.103.024045,PhysRevD.107.044007}. The divergence in that limit is still a mathematical artefact and can be regularised by a proper coordinate redefinition, as explained in \cite{PhysRevD.97.124058}. However, in the short-distance limit, the ambiguity is not present,

{\sl Application.--} Now imagine if future surveys could accurately track the position of large numbers of solar system objects such as asteroids. If the objects were on sufficiently stable orbits, one could imagine estimating the time-dependent perturbation to the apparent position of each object as a function of time and relative to their proper motion in the Solar System Barycentric coordinate system. This would yield a data time stream 
\begin{align}
	d^\phi_I(t)=\delta\phi_I(t)+\zeta^\phi_I(t)\,,\quad d^\theta_I(t)=\delta\theta_I(t)+\zeta^\theta_I(t)\,,
\end{align}
where capital indices, e.g. $I$ indicate each of $N$ objects $I=1,...,N$. We have introduced observational noise terms $\zeta_I(t)$ for each object's time stream. The correlation in the signal component of the time stream will remain invariant under the projection \eqref{dphi_dtheta_def} as chosen for each pair of objects, and one can therefore build estimators to compare to the expected $\cos(\theta)$ and $\cos(2\theta)$ forms.

To carry out a simple estimate of the signal-to-noise ratio (SNR) of the observations, we assume the noise is uncorrelated between the angular measurements in $\theta,\phi$ and also between all objects and that it is stationary, Gaussian, and white. Considering the Fourier transform of the time streams, we then characterize the statistics of the noise in the frequency domain as
\begin{align}
	&\langle \tilde \zeta^\phi_I(f) \tilde \zeta^{\phi\star}_J(f')\rangle\!=\!\langle\tilde \zeta^\theta_I(f) \tilde \zeta^{\theta\star}_J(f')\rangle\!=\!\delta_{IJ}\,\delta(f-f')\,\Sigma\,.
\end{align}
where $\Sigma=\sigma^2\Delta T$, $\sigma$ is the error in each measurement performed by the survey with a cadence $\Delta T$.

For each pair of objects $(I,J)$ with $I\neq J$, we perform the aforementioned rotations and combine the cross-correlators into estimators
\begin{align}
	\mathcal{C}_{IJ}^\phi&\equiv\int df\delta\phi^{\,}_I(f)\delta\phi^\star_J(f)\tilde Q(f)\,,\nonumber\\
	&=\int df\,H(f) Q(f)\cos(\theta_{IJ})\,,\nonumber\\
	\mathcal{C}_{IJ}^\theta&\equiv\int df\delta\theta^{\,}_I(f)\delta\theta^\star_J(f)\tilde Q(f)\,,\nonumber\\
	&=\int df\,H(f) Q(f)\cos(2\theta_{IJ})\,,
\end{align}
where $\theta_{IJ}$ is the angle between the objects and $Q(f)$ a filter function, which we will set in order to maximize the signal-to-noise ratio (SNR)
\begin{align}
	\text{SNR}^2=\sum_{I\neq J}\left[\frac{|\langle\mathcal{C}_{IJ}^\phi\rangle|^2}{\langle|\mathcal{C}_{IJ}^\phi|^2\rangle}+\frac{|\langle\mathcal{C}_{IJ}^\theta\rangle|^2}{\langle|\mathcal{C}_{IJ}^\theta|^2\rangle}\right]\,.
\end{align}
Using the fact that both $\cos(\theta)$ and $\cos(2\theta)$ average to $1/2$, and setting the optimal filter $Q(f)$ in an analogous manner to what has been done in \cite{Mentasti:2023gmg} we obtain
\begin{align}\label{eq_SNR}
	\text{SNR}_{\rm opt}^2&\simeq\sum_{I\neq J}T\int_{1/T}^{1/\Delta T} df \frac{H^2(f)}{\Sigma^2}=\nonumber\\
	&\simeq \frac{T\,N^2}{2\,\sigma^4\Delta T^2}\left(\frac{3H_0^2}{32\pi^3}f_{\text{ref}}^\gamma\right)^2\Omega_0^2\int_{1/T}^{1/\Delta T} df f^{-2(\gamma+3)}\,,
\end{align}
where $T$ is the total time of observations and where we considered a power-law SGWB such that
\begin{align}\label{PSD_to_Omega}
	H(f)=\frac{3H_0^2}{32\pi^3f^3}\Omega_0\left(\frac{f}{f_{\text{ref}}}\right)^{-\gamma}\,,
\end{align} with $H_0=70\,\text{km}\,\text{s}^{-1}\,\text{Mpc}^{-1}$.
We consider two cases for the spectral index $\gamma$ to forecast the sensitivity of the GW energy density measurements $\Omega_{GW}$. Firstly, $\gamma=0$, a scale-invariant spectrum, which cosmological models often predict. Secondly, $\gamma=\frac{13}{3}$, which corresponds to the expected Supermassive Black Hole Binaries (SMBHB) GW background.
For both cases, for $\gamma>-3$ and since $T\gg\Delta T$, a good approximation of \eqref{eq_SNR} is
\begin{align}
	\text{SNR}_{opt}^2&\simeq\frac{N^2}{2\,\sigma^4\Delta T^2}\,\frac{T^{2(\gamma+3)}}{5+2\gamma}\left(\frac{3H_0^2}{32\pi^3}f_{\text{ref}}^\gamma\right)^2\Omega_0^2\,.
\end{align}

The optimally filtered SNR \eqref{eq_SNR} scales differently than in long-distance astrometric observations \cite{GOLAT2021136381}. For backgrounds with $\gamma>-3$, as we forecast here, the SNR scales as $T^{3+\gamma}$. This suggests surveying techniques can be optimised for overall integration and sensitivity rather than cadence.

To yield SNR above unity, the number of objects tracked and/or the astrometric accuracy of the survey would have to be very high given the tentative measurement of the GW power spectrum imposed by the last PTA observations \cite{NANOGrav:2023gor}.

\begin{table}[t!]
	\centering
	\caption{\label{tab:limits}Detection limits for $\Omega_0$}
	\begin{tabular}{llll}
		\hline
		$\sigma$ [mas] & $\gamma$ & $N=1\times 10^5$ & $N=5\times 10^6$ \\
		\hline
		$50.0$ & 0 &$9.6 \times 10^{-5}$ & $1.9 \times 10^{-6}$ \\ [2pt]
		$0.1$ & 0 &$3.8 \times 10^{-10}$ & $7.7 \times 10^{-12}$ \\[2pt]
		$0.01$ & 0 &$3.8 \times 10^{-12}$ & $7.7 \times 10^{-14}$ \\[2pt]
		\hline \\[1pt]
		$50.0$ & 13/3 &$9.6 \times 10^{-5}$ & $1.9 \times 10^{-6}$ \\ [2pt]
		$0.1$ & 13/3 &$3.8 \times 10^{-10}$ & $7.7 \times 10^{-12}$ \\ [2pt]
		$0.01$ & 13/3 &$3.8 \times 10^{-12}$ & $7.7 \times 10^{-14}$ \\
	\end{tabular}
	\begin{center}
		\footnotesize
		\textbf{Note:} Threshold values of $\Omega_0$, as introduced in \eqref{PSD_to_Omega} needed to produce $\text{SNR}>1$ in \eqref{eq_SNR} using asteroid astrometry. We consider two standard spectral indices $\gamma=0$ and 13/3 with $f_{\rm ref}=30$ nHz. The top right value ($N=5\times 10^6$, $\sigma=50$ mas) corresponds to a headline LSST mission specification \cite{2019ApJ...873..111I}, while the first entry of the second row ($N=1\times 10^5$, $\sigma=0.1$ mas) is the upper limit that GAIA could obtain. In the bottom-right corner, we show the ideal case of a survey with high astrometric accuracy and capable of probing a large number of solar system objects ($N=5\times 10^6$, $\sigma=0.01$ mas). A cadency of $\Delta T=3$ days and a total observation time of $T=10$ years are assumed.
	\end{center}
\end{table}

In Table~\ref{tab:limits}, we show threshold values $\Omega_0$ for SNR=1, for two spectral indices, $\gamma=0$ and 13/3, that would be obtained for different survey sizes and astrometric accuracy. For the non-scale invariant case the spectra are normalise at pivot $f_{\rm ref}=30$ nHz. An astrometric accuracy of 50 mas should be achievable by The Rubin observatory. However, its survey covers less than half the sky, and this will limit the number of objects it will track. Ground-based observations are limited to an astrometric accuracy of $\sim 10$ mas by atmospheric distortion \cite{2019ApJ...873..111I}. Space-based surveys will not suffer from this limitation, but achieving Gaia-like astrometric accuracy of $\sim 10 \mu$as for asteroids will require increased sensitivity. However, it is conceivable that future generations of space-based astrometric surveys could track ${\cal O}(10^6)$ asteroids at this with such accuracies. Indeed, future surveys may not be designed with this analysis in mind. However, any survey that tracks large numbers of solar system objects can be exploited to provide complementary limits to those obtained in the long-distance regime. 

Our analysis shows that, for SMBHB backgrounds ($\gamma=13/3$), the sensitivity for solar system objects, short-distance astrometry, could rival that of current PTA measurements that indicate $\Omega_0(f_{\rm ref}=30 \mbox{nHz}) \sim 2\times 10^{-7}$ \cite{NANOGrav:2023gor} for an SMBHB background.

{\sl Discussion.--} Undoubtedly, the observations proposed here will face several other systematic challenges in practice. Asteroids are not well-behaved test masses and are subject to various external forces. The resulting acceleration may dominate noise contributions if the timescales overlap the frequencies being targeted. Asteroids likely suffer from significant intrinsic brightness fluctuations, which may limit the accuracy of astrometric tracking.
The analysis of these observations would also require highly accurate definitions of reference coordinate frames used to translate between Earth-centric observations to SSB coordinates. Notwithstanding these challenges, given the enduring interest in how observations of any SGWB could transform our understanding of astrophysical processes and fundamental physics, it is interesting to consider the prospects of short-distance astrometry.
Our analysis only considers the noise-dominated limit, i.e., when the instrumental noise's average amplitude is greater than the signal. When deviating from this limit (and even for the case of noiseless instruments \cite{PhysRevLett.131.221403}), the effect of cosmic variance \cite{PhysRevD.107.043018} becomes a significant component and needs to be considered to produce realistic forecasts. However, given the constraints in table \ref{tab:limits}, we would not expect to reach this limit soon.

The results of our analysis motivate further detailed studies for the feasibility of this kind of observation using future surveys and for studying how the signal could be estimated using realistic assumptions for the data.

%
%
{\sl Acknowledgments.--} We thank an anonymous referee for their suggestions and insights shared during the review process. G.M. acknowledges support from the Imperial College London Schr\"odinger Scholarship scheme. C.R.C. acknowledges support under a UKRI Consolidated Grant ST/T000791/1. 


\bibliographystyle{apsrev}
\bibliography{refs}
\end{document}